\documentclass[aps,prl,preprint,groupedaddress,nofootinbib]{revtex4-1}

\usepackage{graphicx}

\begin{document}


\title{Analytical Solution for the Size of the Minimum Dominating Set \\ in Complex Networks}


\author{Jose C. Nacher}
\email[]{nacher@is.sci.toho-u.ac.jp}
\affiliation{Department of Information Science, Faculty of Science, Toho University, Miyama 2-2-1, Funabashi, Chiba 274-8510, Japan}

\author{Tomoshiro Ochiai}
\email[]{ochiai@otsuma.ac.jp}
\affiliation{Faculty of Social Information Studies, Otsuma Women's University, 2-7-1 Karakida, Tama-shi, Tokyo 206-8540, Japan
}

\date{\today} 

\begin{abstract}
Domination is the fastest-growing field within graph theory with a profound diversity and impact in 
real-world applications, such as the recent breakthrough approach that identifies optimized subsets of 
proteins enriched with cancer-related genes. Despite its conceptual simplicity, domination is a classical 
NP-complete decision problem which makes analytical solutions elusive and poses difficulties to design optimization 
algorithms for finding a dominating set of minimum cardinality in a large network. 
Here we derive for the first time an approximate analytical solution for the density of the minimum dominating set (MDS) 
by using a combination of cavity method and Ultra-Discretization (UD) procedure. The derived equation allows us 
to compute the size of MDS by only using as an input the information of the degree distribution of a given network.
\end{abstract}


\maketitle


\section{Introduction}
The research on complex networks in diverse fields \cite{newman, caldarelli}, based on applied graph theory combined 
with computational and statistical physics methods, has experienced a spectacular growth in recent years and has
led to the discovery of ubiquitous patterns called scale-free networks \cite{barabasi}, 
unexpected dynamic behavior \cite{vespignani}, 
robustness and vulnerability features \cite{havlin1, reka, motter}, and applications in natural and social
complex systems \cite{newman, caldarelli}. On the other hand, domination is an important problem in graph theory which has rich 
variants, such as independence, covering and matching \cite{teresa}. The mathematical and computational studies on domination 
have led to abundant applications in disparate fields such as mobile computing and computer communication networks \cite{sto}, 
design of large parallel and distributed systems \cite{swapped}, analysis of large social networks \cite{social1, social2, social3}, 
computational biology and biomedical analysis  \cite{vazquez} and discrete algorithms research \cite{teresa}.

More recently, the Minimum Dominating Set (MDS) has drawn the attention
researchers to controllability in complex networks \cite{nacher1, nacher2, nacher3}, 
to investigate observability in power-grid \cite{power} and to identify an optimized subset of proteins enriched 
with essential, cancer-related and virus-targeted genes in 
protein networks \cite{wuchty}.  The size of MDS was also investigated by extensively analysing several types 
of artificial scale-free networks using a greedy algorithm in \cite{molnar}. The problem to design complex networks that are structurally 
robust has also recently been
investigated using the MDS approach \cite{nacher4, molnar2}.

Despite its conceptual simplicity, the MDS is a classical NP-complete decision problem in computational complexity theory \cite{johnson}. 
Therefore, 
it is believed that there is no a theoretically efficient (i.e., polynomial time) algorithm that finds an exact smallest dominating set for a given graph. 
It is worth noticing that although it is an NP-hard
problem, recent results have shown that we can use Integer Linear Programming (ILP) to find optimal solution \cite{nacher1, wuchty, nacher4}. Moreover, for 
specific types of graph such a tree
(i.e., no loops) and even partial-$k$-tree, it can be solved using Dynamic Programing (DP) in polynomial time \cite{akutsu}. 

Especially, the density (or fraction) of MDS, defined as the size of MDS versus the total number of nodes in a network, is important to 
estimate the cost-efficient network deployment of controllers. It is less expensive to have ten power plants operating as controllers in a 
large network of 1,000 sites than one hundred. 
Similarly, acting on two protein targets via drug interactions 
is always better than acting on ten protein targets to minimize adverse side effects.
Although we can use an ILP method for obtaining an MDS and the density of MDS, we do not have an analytic solution for the density of MDS. 
Note that even though optimal solution for MDS can be found using ILP method, this kind of technique is very generic and operates as 
a {\it black box} so that its systematic application does not provide us any knowledge about the details of the particular problem under study. Moreover, from a physical point of view, analytic solutions always enable us to have a deeper understanding of the problem by examining dependences with other variables. Here, we derive for the first time an analytical solution for an MDS by using cavity method.

Cavity method is a well-known methodology developed by physicists working on statistical mechanics (e.g. spin glasses \cite{spin}). However, this technique 
has also been extended and applied to non-physical systems, including network theory in which nodes can be abstracted to some extent. For 
example, applications include the analysis of random combinatorial problems such as weighted matching \cite{mezard1}, the vertex
cover \cite{hart} and the travelling salesman problem \cite{mezard2}. Moreover,
recently cavity method was used to determine the number of matchings in random graphs \cite{mezard_last}, the 
maximal independent sets \cite{dalla}, the random set packing \cite{indian} and the minimum weight Steiner tree 
\cite{steiner} and 
controllability using maximum matching \cite{liu}, extending the work done in \cite{mezard_last} 
for undirected networks to directed networks.\footnote{Note added: After finishing our paper, we became aware of a recent paper done independently by Zhao et al.\cite{Zhao}, in which the statistical properties of the MDS were studied. Although the starting point of the cavity analysis is similar, in their results they estimate ths size of MDS by population dynamics simulations. In contrast, as shown later, we derive analytical solution for the size of the MDS.
}
 It is also well-known that cavity method computed at zero temperature limit has been 
applied to networks and used in many works, which have led to derive elegant analytical formulas \cite{new1, new2, new3}.

On the other hand, the Ultra Discretization (UD) procedure has been developed mainly in the field of soliton theory and cellular automaton (CA) theory \cite{UD1,UD2,UD3,UD4,UDours}. Although 
the common discretization process discretizes independent variables, the UD procedure can discretize dependent variables. As a result, both 
independent and dependent variables become discrete variables. In other words, UD transforms discretized equations into their corresponding 
UD equations. In soliton theory, we can obtain the cellular automaton from the corresponding discretized soliton theory after UD procedure. In \cite{UD4}, it is shown that the corresponding integrable CA is obtained from Korteweg–de Vries (KdV)
 soliton equation via Lotka-Volterra equation, using UD procedure.
The key point of UD is that the UD procedure can be applied only in the case that the original 
discretized equation does not have any minus operators.

By combining the cavity method and UD procedure, we solved for the first time the density 
of MDS problem analytically for any complex network. By cavity method and UD, we derived a combinatorial 
expression that allows us to identify the density of MDS
by only using the degree distribution of the network, which is our main result. We then compare the 
results of the derived expression for density of MDS with the ILP solutions for regular, random and scale-free networks 
showing a fair agreement. Moreover, a simple formula is derived for random networks at the large degree limit.

\section{Theoretical results}
\subsection{The Hamiltonian of a dominating set}
\begin{figure}[htbp] 
  	\includegraphics[scale=0.7]{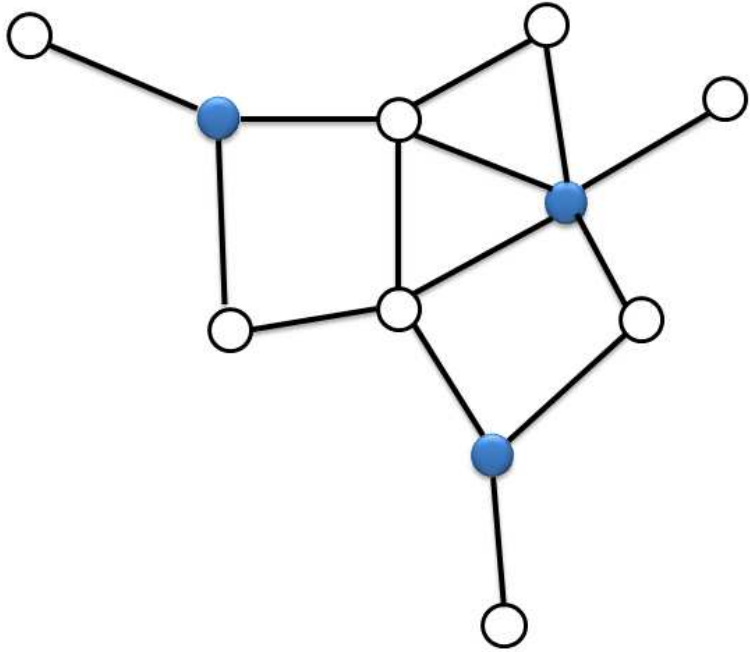}
	\caption{Illustration of an MDS. Filled nodes belong to the MDS. }
	\label{Fig: Example of MDS}
\end{figure}

A graph $G=(V,E)$ is a set of nodes $V$ and edges $E$. A Dominant Set (DS) $S$ is defined to be a subset of $V$, where each node $i\in V$ 
belongs to $S$ or adjacent to an element of $S$. For each node $i\in V$, we define a binary function $\sigma_i$ to be $+1$ if $i\in S$, 
otherwise $0$. We call the node $i$ occupied (resp. empty) if $\sigma_i$ is $+1$ (resp. $0$). The set of the adjacent nodes to node $i$ 
is denoted by $\partial i$. We can then see that the constraint to define a DS is as follows:
\begin{eqnarray}
	\sigma_i+\sum_{j\in \partial i}\sigma_j \ge1.
\end{eqnarray}
 For each node $i\in V$, we define a binary function $I_i$ to be $+1$ if and only if $\sigma_i+\sum_{j\in \partial i}\sigma_j \ge1$, 
 otherwise 0. A set $S$ is an Minimum Dominating Set (MDS) if the size is smallest among all dominating sets (see Fig. \ref{Fig: Example of MDS}).  

Let us consider the following Hamiltonian function:
\begin{eqnarray}
H(\sigma)=\sum_{i\in V}\sigma_i, \label{eqn: hamiltonian}
\end{eqnarray}
where $\sigma=\{ \sigma_i | i\in V\}$. Then the partition function is given by
\begin{eqnarray}
Z(\beta)=\sum_{\sigma} (\prod_{i\in V}I_i) \exp(-\beta H(\sigma)),
\end{eqnarray}
where the summation is taken over all configurations of $\sigma$ and $\beta$ is the inverse temperature.

\subsection{Cavity method analysis}
\begin{figure}[htbp] 
  	\includegraphics[scale=0.7]{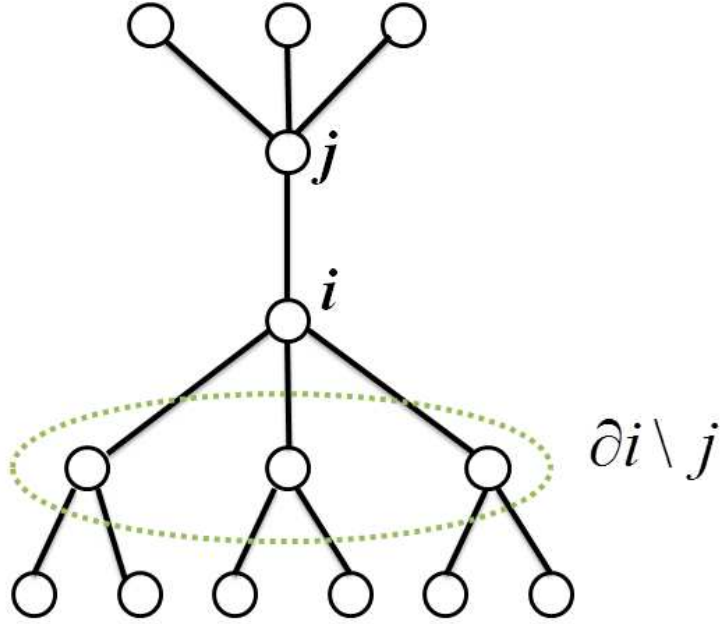}
	\caption{An example of a tree graph. The highlighted nodes $\partial i \backslash j$ indicate the set of nodes adjacent to $i$ except for node $j$.}
	\label{Fig: ij graph}
\end{figure}

In what follows, we apply the cavity method to derive the analytic formula for the density of MDS in complex networks. Note that the 
cavity approach has been used in a similar way to solve the maximal independent set problem \cite{dalla}. First, we assume the graph $G$ 
has tree structure and let $\partial i \backslash j$ be the set of $\partial i$ except for node $j$ (see Fig. \ref{Fig: ij graph}). Note also that the minimum degree of the graph should be two, otherwise some of $\partial i \backslash j$ do not exist. Then, we can write $\sigma_{i \to j}=\{\sigma_k | k \in\partial i \backslash j \}$.  Next, let $\nu_{i \to j}(\sigma_i, \sigma_{i \to j})$ be the probability that the node $i$ and $\partial i\backslash j$ take value $\sigma_i$ and $\sigma_{i \to j}$, respectively, and constraints $I_i$ and $I_j$ are not included:
\begin{eqnarray}
\nu_{i \to j}(\sigma_i, \sigma_{i \to j})=\frac{1}{Z_{i\to j}} \sum_{\sigma^\prime} (\prod_{\alpha \neq i, j} I_\alpha) \exp(-\beta H(\sigma)), \label{eqn: nu}
\end{eqnarray}
where the first summation is taken over all configurations of $\sigma^\prime=\{ \sigma_\beta | (\beta \neq i) \wedge (\beta \notin \partial i\backslash j) \wedge (\beta \in V)\}$. 
The product for the constraint $I_\alpha$ is taken over all nodes of $V$ except for $i$, $j$ (i.e. $\alpha \in V$ and $\alpha \neq i, j$) and $Z_{i\to j}$ is the normalization constant given by
\begin{eqnarray}
Z_{i\to j}&=& \sum_{\sigma_i}\sum_{\sigma_{i \to j}} \sum_{\sigma^\prime} (\prod_{\alpha \neq i, j} I_\alpha) \exp(-\beta H(\sigma)) \nonumber\\
&=& \sum_{\sigma} (\prod_{\alpha \neq i, j} I_\alpha) \exp(-\beta H(\sigma)).
\end{eqnarray}

\begin{figure}[htbp] 
  	\includegraphics[scale=0.7]{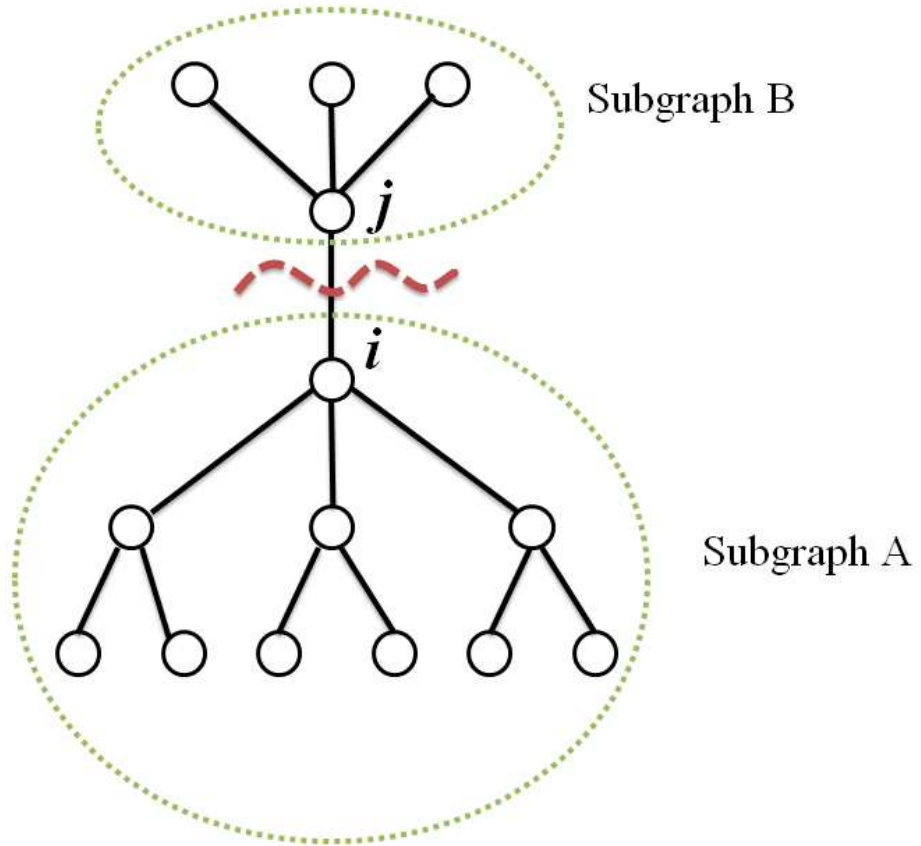}
	\caption{Two subgraphs ($A$ and $B$) are obtained by cutting an edge ($i$, $j$). Note that subgraph $A$ and $B$ includes nodes $i$ and $j$, respectively. }
	\label{Fig: cut into subgraph}
\end{figure}

Eq. (\ref{eqn: nu}) can be written in a different way, which will be easier to compute. As shown in Fig. \ref{Fig: cut into subgraph}, we divide a graph into two subgraphs ($A$ and $B$) by cutting the edge ($i$, $j$) between node $i$ and $j$. Let $A$ be the set of all the nodes which belongs to the subgraph including $i$, and $B$  be the set of all the nodes which belongs to the subgraph including $j$. Then, Eq. (\ref{eqn: nu}) can be transformed into
\begin{eqnarray}
\nu_{i \to j}(\sigma_i, \sigma_{i \to j})=\frac{1}{Z^\prime_{i\to j}} \sum_{\sigma^\prime_A} (\prod_{\alpha \neq i} I_\alpha) \exp(-\beta \sum_{\gamma\in A}\sigma_\gamma),
\end{eqnarray}
where the first summation is taken over all configurations of $\sigma^\prime_A=\{ \sigma_\beta | (\beta \neq i) \wedge (\beta \notin \partial i\backslash j) \wedge (\beta \in A)\}$. The product for the constraint $I_\alpha$ is taken over all nodes of $A$ except $i$ (i.e. $\alpha \in A$ and $\alpha \neq i$) and $Z^\prime_{i\to j}$ is a normalization constant given by
\begin{eqnarray}
Z^\prime_{i\to j}&=& \sum_{\sigma_i}\sum_{\sigma_{i \to j}} \sum_{\sigma^\prime_A} (\prod_{\alpha \neq i} I_\alpha) \exp(-\beta \sum_{\gamma\in A}\sigma_\gamma)\nonumber\\
&=& \sum_{\sigma_A} (\prod_{\alpha \neq i} I_\alpha) \exp(-\beta \sum_{\gamma\in A}\sigma_\gamma),
\end{eqnarray}
where the last summation is taken over all configurations of $\sigma_A=\{ \sigma_\beta | (\beta \in A)\}$. 

Then the following exact recursive equation can be derived:
\begin{eqnarray}
\nu_{i \to j}(\sigma_i, \sigma_{i \to j})\propto e^{-\beta \sigma_i}
\prod_{k \in \partial i\backslash j}
\sum_{\sigma_{k\to i}} I_k \nu_{k \to i}(\sigma_k, \sigma_{k \to i}).
\end{eqnarray}
This equation can be iteratively solved with its corresponding normalization constant.


Let $\bar{\nu}(\sigma_i,m)$ be the summation of $\nu_{i \to j}(\sigma_i, \sigma_{i \to j})$, where the number of occupied neighbors $\sigma_{i \to j}$ is $m$: 
\begin{eqnarray}
\bar{\nu}_{i \to j}(\sigma_i, m) = \sum_{\sum_{\alpha \in \partial i \backslash j}\sigma_\alpha=m} \nu_{i \to j}(\sigma_i, \sigma_{i \to j}).
\end{eqnarray}

Then, let $r_1^{i \to j}$ be the probability that the node $i$ is occupied in the cavity graph. Let $r_{00}^{i \to j}$ be the probability that the node $i$ and all the neighbors $\sigma_{i \to j}$ are empty. Let $r_0^{i \to j}$ be the probability that the node $i$ is empty and  at least one of the neighbors $\sigma_{i \to j}$ are occupied. Then, we can write:
\begin{eqnarray}
r_1^{i \to j} &=& \sum_{m=0}^{k_i-1} \bar{\nu}_{i \to j}(1, m),\\
r_{00}^{i \to j} &=& \bar{\nu}_{i \to j}(0, 0),\\
r_0^{i \to j} &=& \sum_{m=1}^{k_i-1} \bar{\nu}_{i \to j}(0, m),
\end{eqnarray}
where $k_i$ is the degree of node $i$. Here we note that $r_1^{i \to j}+r_{00}^{i \to j}+r_0^{i \to j}=1$, because of probability normalization.

The above definitions lead to the following iterative equations:
\begin{eqnarray}
r_1^{i \to j} &=& \frac{1}{N} e^{-\beta}, \label{iterative equation1}\\
r_{00}^{i \to j} &=& \frac{1}{N} \prod_{k\in \partial i \backslash j} r_0^{k \to i}, \label{iterative equation2}\\
r_0^{i \to j} &=& \frac{1}{N} \Bigl{\{} \prod_{k\in \partial i \backslash j}(1-r_{00}^{k \to i}) - \prod_{k\in \partial i \backslash j} r_0^{k \to i} \Bigr{\}}, \label{iterative equation3}
\end{eqnarray}
where $N$ is the normalization constant given by
\begin{eqnarray}
N &=& e^{-\beta} +  \prod_{k\in \partial i \backslash j}(1-r_{00}^{k \to i}) .
\end{eqnarray}

\subsection{Ultra-Discretization (UD) procedure}

We note that until now we are considering the problem of an DS, which is treated as finite temperature problem ($\beta$ is finite) in the context of statistical mechanics. To address the minimum dominating set (MDS) problem,
we have to consider the zero-temperature limit ($\beta \to \infty$) which gives the ground energy state of the Hamiltonian shown in Eq. (\ref{eqn: hamiltonian}). To solve the equations associated to the zero-temperature limit, we can use an Ultra-Discretization (UD) procedure \cite{UD4}. 

\begin{table}[htb]
  \begin{tabular}{|l|c|c|}\hline
	Equation & independent variable & dependent variable\\ \hline \hline
    ($i$) Continuous equation for $x(t)$ & $t$: real value & $x(t)$:  real value \\ \hline
    ($ii$) Discretized equation  for $x_n$ &  $n$:  integer & $x_n$: real value\\ \hline
    ($iii$) Ultra- Discretized (UD) equation for $X_n$ & $n$: integer & $X_n$: integer \\ \hline
  \end{tabular}
	\caption{Discretization procedure changes the equation type from ($i$) to ($ii$) and Ultra-Discretization procedure transforms the equation type from ($ii$) to ($iii$).}
\end{table}

Discretization is a well-known method to discretize independent variable in continuous theory such as differential equation (see Table 1 from ($i$) to ($ii$)). UD can go further and aims to discretize the dependent variable in the discretized equation (from ($ii$) to ($iii$)). As a result, in the transformed UD equation ($iii$), 
both independent and dependent variables are discretized (i.e. both are integer variables). The key formula of UD, which transforms the equation type from ($ii$) to ($iii$)  is
\begin{eqnarray}
\lim_{\beta \to \infty}\frac{1}{\beta}\log(e^{\beta X}+e^{\beta Y})=\max(X,Y). 
\end{eqnarray}
When the discretized equation ($ii$) has no subtraction operator, we can transform it into the corresponding UD equation ($iii$) by replacing $x=e^{\beta X}$ and  $y=e^{\beta Y}$ and taking limit $\beta \to \infty$. Here, $x$ and $y$ belong to discretized equation ($ii$) and $X$ and $Y$ belong to UD equation ($iii$).  The operator in the original discretized equation ($ii$) is transformed in the UD equation ($iii$) as follows:
\begin{eqnarray}
&&x\times y \to X+Y,\\
&&x / y \to X-Y,\\
&&x + y \to \max(X,Y).
\end{eqnarray}
Here we remark that in order to obtain the UD equation ($iii$), it is necessary to be able to remove any minus operator in the original discretized equation ($ii$).

In order to ultra-discretize the previous Eqs. (\ref{iterative equation1}), (\ref{iterative equation2}) and (\ref{iterative equation3}), we make them consist of only plus, multiplication and division operators, avoiding minus operators. By simple computation, we have
\begin{eqnarray}
\prod_{k\in \partial i \backslash j}(1-r_{00}^{k \to i}) = \prod_{k\in \partial i \backslash j}(r_{0}^{k \to i}+r_{1}^{k \to i}) \label{eqn: minus avoiding 1}
\end{eqnarray}
and
\begin{eqnarray}
\prod_{k\in \partial i \backslash j}(1-r_{00}^{k \to i}) - \prod_{k\in \partial i \backslash j} r_0^{k \to i} = \sum_{1 \le m_1 + \cdots + m_n \le n} r_{m_1}^{k_1 \to i} r_{m_2}^{k_2 \to i} \cdots r_{m_n}^{k_n \to i}, \label{eqn: minus avoiding 2}
\end{eqnarray}
where $n$ is the number of elements of $\partial i \backslash j$, and $m_p = 0~\mbox{or}~ 1$ $(p=1,2,3,\cdots,n)$.

By inserting Eqs. (\ref{eqn: minus avoiding 1}) and (\ref{eqn: minus avoiding 2}) into Eqs. (\ref{iterative equation1}), (\ref{iterative equation2}) and (\ref{iterative equation3}), we have
\begin{eqnarray}
r_1^{i \to j} &=& \frac{e^{-\beta}}{e^{-\beta} +  \prod_{k\in \partial i \backslash j}(r_{0}^{k \to i}+r_{1}^{k \to i}) }, \label{eqn: iterative equation no minus 1}\\
r_{00}^{i \to j} &=& \frac{\prod_{k\in \partial i \backslash j} r_0^{k \to i}}{e^{-\beta} +  \prod_{k\in \partial i \backslash j}(r_{0}^{k \to i}+r_{1}^{k \to i})}, \label{eqn: iterative equation no minus 2}\\
r_0^{i \to j} &=& \frac{\sum_{1 \le m_1 + \cdots + m_n \le n} r_{m_1}^{k_1 \to i} r_{m_2}^{k_2 \to i} \cdots r_{m_n}^{k_n \to i} }{e^{-\beta} +  \prod_{k\in \partial i \backslash j}(r_{0}^{k \to i}+r_{1}^{k \to i})} .\label{eqn: iterative equation no minus 3}
\end{eqnarray}
Here we remark that the above derived equations consist only of three operators (plus, multiplication and division), which is suitable for the ultra-discretization.  

Here, we then replace the following three variables as follows:
\begin{eqnarray}
r_1^{i \to j} &=& e^{\beta R_1^{i \to j} }, \label{eqn:UD variable 1}\\
r_{00}^{i \to j} &=& e^{\beta R_{00}^{i \to j}},  \label{eqn:UD variable 2}\\
r_0^{i \to j} &=& e^{\beta R_{0}^{i \to j}},\label{eqn:UD variable 3}
\end{eqnarray}
where $R_1^{i \to j}$, $R_{00}^{i \to j}$ and $R_{0}^{i \to j}$ are variables in UD system.

After inserting (\ref{eqn:UD variable 1}), (\ref{eqn:UD variable 2}), (\ref{eqn:UD variable 3}) into (\ref{eqn: iterative equation no minus 1}),  (\ref{eqn: iterative equation no minus 2}), (\ref{eqn: iterative equation no minus 3}) and taking zero-temperature limit $\beta \to \infty$, Eqs. (\ref{eqn: iterative equation no minus 1}),  (\ref{eqn: iterative equation no minus 2}) and  (\ref{eqn: iterative equation no minus 3}) are transformed into
\begin{eqnarray}
R_1^{i \to j} &=& -1 - \max(-1, \sum_{k\in \partial i \backslash j} \max(R_{0}^{k \to i}, R_{1}^{k \to i})), \label{eqn:UD equation 1}\\
R_{00}^{i \to j} &=& \sum_{k\in \partial i / j} R_0^{k \to i} - \max(-1,  \sum_{k\in \partial i \backslash j} \max(R_{0}^{k \to i}, R_{1}^{k \to i})), \label{eqn:UD equation 2}\\
R_0^{i \to j} &=& \max_{1 \le m_1 + \cdots + m_n \le n} (R_{m_1}^{k_1 \to i} + R_{m_2}^{k_2 \to i} + \cdots + R_{m_n}^{k_n \to i}) -  \max(-1,  \sum_{k\in \partial i \backslash j} \max(R_{0}^{k \to i}, R_{1}^{k \to i})). \nonumber\\ \label{eqn:UD equation 3}
\end{eqnarray}

Eq. (\ref{eqn: iterative equation no minus 1}),  (\ref{eqn: iterative equation no minus 2}), (\ref{eqn: iterative equation no minus 3}) (or Eq. (\ref{eqn:UD equation 1}), (\ref{eqn:UD equation 2}), (\ref{eqn:UD equation 3}))
are very difficult to solve, because every edge direction has three variables and three associated equations. Therefore, in order to avoid this high complexity, we use a coarse-grained method.

Let $p_k$ be the degree distribution of the network which gives the probability to find nodes with degree $k$ and $\langle k \rangle=\sum_k k p_k$ be the average degree of the network. The excess degree distribution is given by $q_k= (k+1)p_{k+1}/{\langle k \rangle}$, that is the probability to find that a neighbor node has degree k.

Let us assume that the network is enough large. Each $r_1^{i \to j}$ and $r_0^{i \to j}$ values are assigned to the edge direction from $i$ to $j$. The opposite edge direction gets the values from $r_1^{j \to i}$ and $r_0^{j \to i}$. $r_{00}^{i \to j}$ is not independent variable because of the normalization. Let $P(r_1,r_0)$ be the probability density function of $r_1^{i \to j}$, $r_0^{i \to j}$.

Then we have the coarse-grained equation (cavity mean field equation) for (\ref{eqn: iterative equation no minus 1}) and (\ref{eqn: iterative equation no minus 3})  as follows:
\begin{eqnarray}
P(r_1,r_0)&=&\sum_{k=1}^\infty q_k\int\prod_{l=1}^k dr_1^l dr_0^l P(r_1^l,r_0^l)\nonumber\\
&&\times \delta(r_1 - \frac{e^{-\beta}}{e^{-\beta} +  \prod_{l=1}^k(r_{0}^l+r_{1}^l) })\nonumber\\
&&\times \delta(r_0 - \frac{\sum_{1 \le m_1 + \cdots + m_k \le k} r_{m_1}^{1} r_{m_2}^{2} \cdots r_{m_k}^{k} }{e^{-\beta} +  \prod_{l=1}^k(r_{0}^{l}+r_{1}^{l})}). \label{eqn: coarse-grained equation}
\end{eqnarray}

We transform the probability density function by variables transformation (\ref{eqn:UD variable 1}) and (\ref{eqn:UD variable 3}) as follows:
\begin{eqnarray}
P(r_1,r_0)dr_1dr_0=\bar{P}(R_1,R_0)dR_1dR_0.
\end{eqnarray}
Then, we have
\begin{eqnarray}
\bar{P}(R_1,R_0)&=&\sum_{k=1}^\infty q_k\int\prod_{l=1}^k dR_1^l dR_0^l \bar{P}(R_1^l,R_0^l)\nonumber\\
&&\times \delta(r_1 - \frac{1}{\beta}\log \frac{e^{-\beta}}{e^{-\beta} +  \prod_{l=1}^k(e^{\beta R_{0}^l}+e^{\beta R_{1}^l}) })\nonumber\\
&&\times \delta(r_0 - \frac{1}{\beta}\log \frac{\sum_{1 \le m_1 + \cdots + m_k \le k} e^{\beta R_{m_1}^{1}} e^{\beta R_{m_2}^{2}} \cdots e^{\beta R_{m_k}^{k}} }{e^{-\beta} +  \prod_{l=1}^k(e^{\beta R_{0}^l}+e^{\beta R_{1}^l})}).
\end{eqnarray}

Taking UD limit (zero temperature) $\beta \to \infty$, we have the ultra-discretization version of cavity equation:
\begin{eqnarray}
\bar{P}(R_1,R_0)&=&\sum_{k=1}^\infty q_k\int\prod_{l=1}^k dR_1^l dR_0^l \bar{P}(R_1^l,R_0^l) \nonumber\\
&&\times \delta(R_1 +1 + \max(-1, \sum_{l=1}^k \max(R_{0}^l, R_{1}^l)) ) \nonumber\\
&&\times \delta(R_0 - \max_{1 \le m_1 + \cdots + m_k \le k} (R_{m_1}^1 + R_{m_2}^2 + \cdots + R_{m_k}^k) + 
\max(-1, \sum_{l=1}^k \max(R_{0}^l, R_{1}^l))  ). \nonumber\\
\label{eqn: cavity equation UD}
\end{eqnarray}

Next, we will solve this equation. Eqs. (\ref{eqn:UD equation 1}), (\ref{eqn:UD equation 2}), (\ref{eqn:UD equation 3}) imply that $R_1$ and $R_0$ takes integer values, since $R_1$ and $R_0$ at the boundary of network takes integer values.  Furthermore, considering the probability conservation, Eqs. (\ref{eqn:UD variable 1}), (\ref{eqn:UD variable 3}) imply that $R_1$ and $R_0$ takes value 0 or negative integer. By considering Eq. (\ref{eqn:UD equation 1}), firstly, we can see that $R_1^{i \to j}$ takes value only $0$ or $-1$. Secondly, if $R_1^{i \to j}$ is $-1$, then for each $k\in \partial i \backslash j$, one of $R_{0}^{k \to i}$ or $R_{1}^{k \to i}$ takes value $0$. In this case ($R_1^{i \to j}=-1$), from Eq. (\ref{eqn:UD equation 3}),  
$R_0^{i \to j}$ takes value $0$ or $-1$. Therefore, we can set the distribution $\bar{P}(R_1,R_0)$ as follows:
\begin{eqnarray}
\bar{P}(R_1,R_0)=a\delta(R_1 + 1)\delta(R_0) + b \delta(R_1+1)\delta(R_0+1)+\sum_{n=0}^\infty c_n \delta(R_1)\delta(R_0 + n), \label{eqn: anzats}
\end{eqnarray}
where $a+b+\sum_{n=0}^\infty c_n=1$.

Inserting (\ref{eqn: anzats}) into the cavity equation (\ref{eqn: cavity equation UD}), we have
\begin{eqnarray}
a &=& \sum_{k=1}^\infty q_k ((1-b)^k - a^k), \label{eqn: parameter eq 1}\\
b &=& \sum_{k=1}^\infty q_k a^k, \label{eqn: parameter eq 2}\\
c_{m-1} &=& \sum_{k=m}^\infty q_k ({}_kC_m)  b^m(1-b)^{k-m}~~~~~(m=1,2,3,\cdots). \label{eqn: parameter eq 3}
\end{eqnarray}
Once the degree distribution $p_k$ is given, we can determine $a$, $b$, $c_n$ $(n=0,1,2,\cdots)$.

Here, we need the average energy of the Hamiltonian (\ref{eqn: hamiltonian}), 
which can be identified as the average number of all DS configurations. 
After taking zero temperature limit $\beta \to \infty$, we will obtain the analytical 
formula for the density of MDS.


The average energy can be computed by $<H>=-\frac{\partial}{\partial \beta}\log Z$, where the partition function is shown in Supplementary Information. A generic expression for density is given by
\begin{eqnarray}
\rho&=&\frac{<H>}{|V|} \nonumber\\
&=&\sum_{k=2}^\infty p_k\int\prod_{l=1}^k dr_1^l dr_0^l P(r_1^l,r_0^l)\frac{e^{-\beta}}{e^{-\beta}+\prod_{l=1}^k(r_0^l + r_{00}^l) - \prod_{l=1}^k r_{0}^l }.
\end{eqnarray}
This equation, however, cannot be ultra-discretized because of the minus operator, and without being ultra-discretized, analytical solution cannot be obtained. Therefore, we avoid minus operator by inserting Eqs. (21) and (22), which leads to the following result:
\begin{eqnarray}
\rho&=&\sum_{k=2}^\infty p_k\int\prod_{l=1}^k dr_1^l dr_0^l P(r_1^l,r_0^l)\frac{e^{-\beta}}{e^{-\beta}+\sum_{1\le m_1 + \cdots + m_k \le k}(r_{m_1}^1\cdots r_{m_k}^k)}.
\end{eqnarray}


After taking zero temperature limit $\beta \to \infty$, we obtain 
\begin{eqnarray}
\rho&=&\lim_{\beta \to \infty}\sum_{k=2}^\infty p_k\int\prod_{l=1}^k dR_1^l dR_0^l P(R_1^l,R_0^l)\frac{e^{-\beta}}{e^{-\beta}+\sum_{1 \le m_1 + \cdots + m_k \le k} e^{\beta(R_{m_1}^1 + R_{m_2}^2 + \cdots + R_{m_k}^k)} }
\end{eqnarray}

By inserting (\ref{eqn: anzats}) into this equation,  we finally obtain
\begin{eqnarray}
\rho&=&\sum_{k=2}^\infty p_{k}\Bigl[1-(1-b)^k-kb(1-b)^{k-1}+a^{k}\frac{1}{k+1} \nonumber\\
&&+\sum_{p+q+r=k-1}\frac{k!}{p!r!s!}a^pb(c_0)^r(1-b-c_0)^{s} \frac{1}{1+2^{r+1}} \nonumber\\
&&+\sum_{p+r=k-1}\frac{k!}{p!r!}a^pb(c_0)^r(\frac{1}{1+2^{r+1}}-\frac{1}{2^{r+1}})\Bigr]. \label{eqn: analytic formula1}
\end{eqnarray}
More simply, we transform the previous equation into
\begin{eqnarray}
\rho&=&\sum_{k=2}^\infty p_k\Bigl[1-(1-b)^k-kb(1-b)^{k-1}+a^k\frac{1}{k+1} \nonumber\\
&&+\sum_{r=0}^{k-1}\frac{k!}{(k-r-1)!r!}bc_0^r\times \bigl\{(1-b-c_0)^{k-r-1} \frac{1}{1+2^{r+1}} - 
a^{k-r-1}(\frac{1}{1+2^{r+1}}-\frac{1}{2^{r+1}}\bigr\}\Bigr]. \nonumber\\\label{eqn: analytic formula2}
\end{eqnarray}
Eq. (\ref{eqn: analytic formula2}) is our main result. Using this equation, we can analytically compute the density of MDS from any degree distribution $p_k$. 
 More concretely, we can summarize the procedure as follows. Once the degree density 
function $p_k$ is given, we can easily compute the excess degree density function $q_k$. Then, by solving 
(\ref{eqn: parameter eq 1}), (\ref{eqn: parameter eq 2}) and (\ref{eqn: parameter eq 3}), we 
obtain $a$, $b$, $c_n$ $(n=0,1,2,\cdots)$. Finally, inserting the value $a$, $b$, $c_n$ $(n=0,1,2,\cdots)$ 
into the above equation (\ref{eqn: analytic formula2}), we obtain the density of MDS.

We note that when we consider random network with average degree $z$ as a special case,  we can derive the simple expression $\rho = 1/z$ by using large average degree limit approximation ($z \to \infty$) (See supplementary Information).

\section{Computational results}
\begin{figure}[htbp] 
  	\includegraphics[scale=0.5]{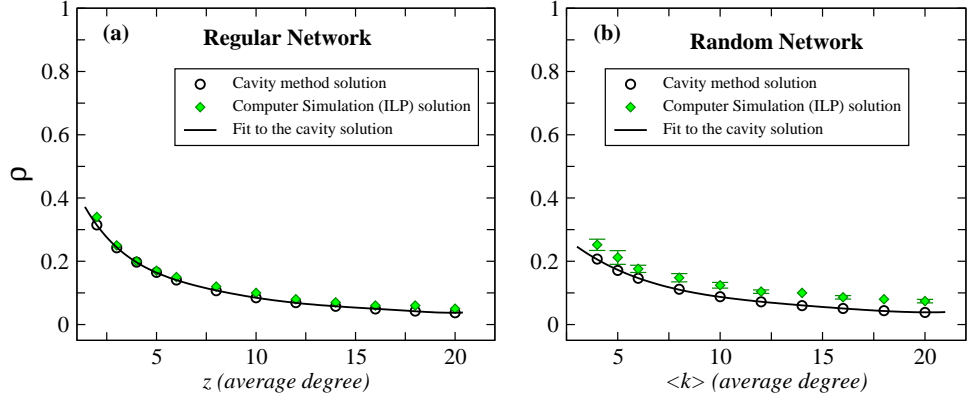}
	\caption{ Comparison of the cavity method and ILP results for the MDS density $\rho$ in regular (a) and random (b) networks constructed with a variety of average degree values.  The results are averaged using five network samples. For cavity method
the standard deviation (SD) is smaller than symbol. For ILP the SD is shown for each symbol.   
}
	\label{fig: random regular}
\end{figure}

Here we performed computer simulations to examine the theoretical results (\ref{eqn: analytic formula2}) obtained using cavity method.
First, we consider regular and random networks constructed with a variety of average degree values (see Fig. \ref{fig: random regular} (a) and (b)).  The results show that cavity method predictions are in excellent agreement with ILP solutions.

\begin{figure}[htbp] 
  	\includegraphics[scale=0.5]{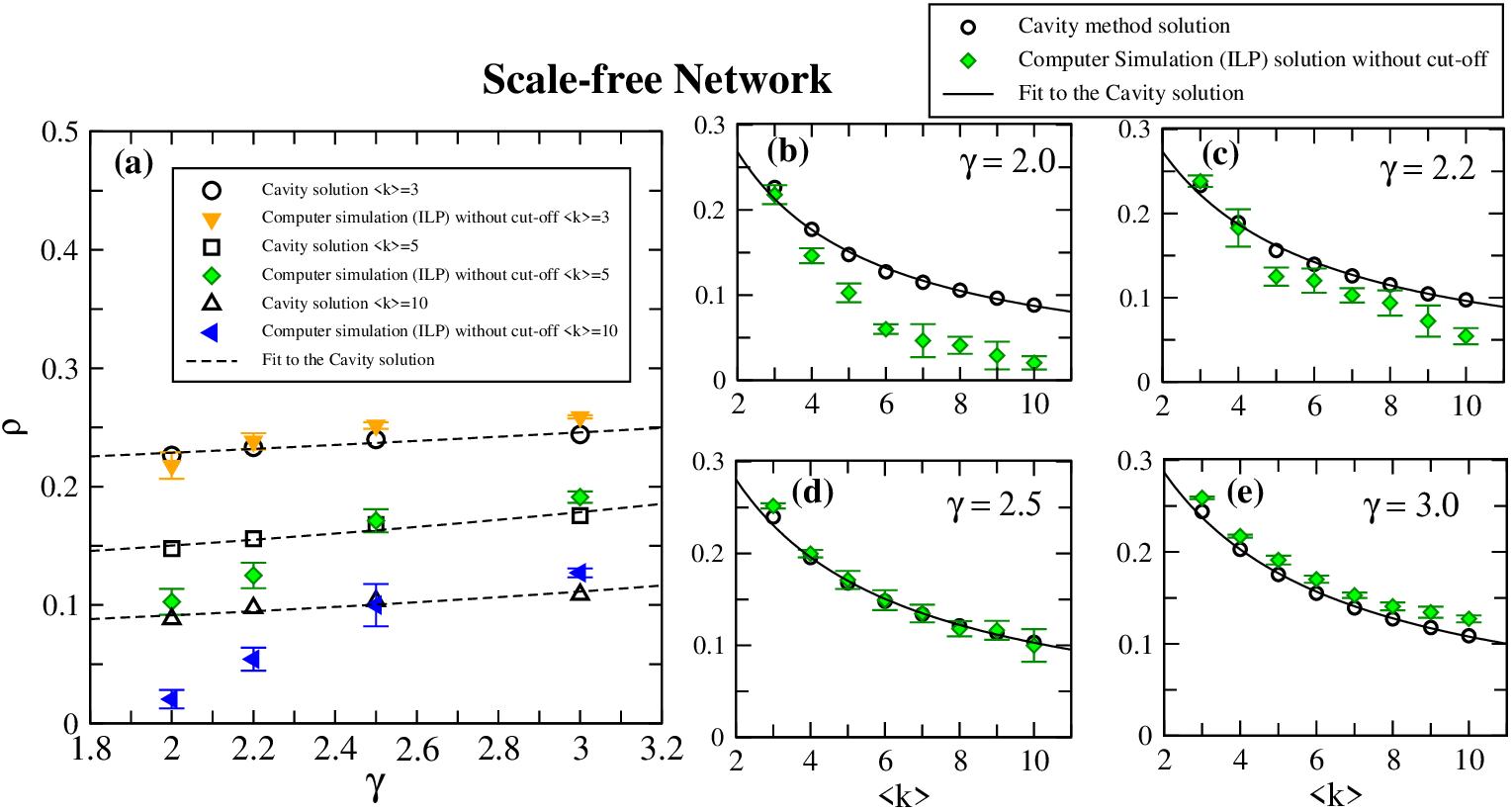}
	\caption{Comparison of the cavity method and ILP results for the MDS density $\rho$ in samples of synthetic scale-free networks generated without structural cut-off $k_c=N-1$. $\rho$ as a function of the scaling exponent $\gamma$ (a) and average degree $<k>$ (b-d) are shown in figure. The results are averaged using five network samples with a size of $N=5000$ nodes. For cavity method
the standard deviation (SD) is smaller than symbol. For ILP the SD is shown for each symbol.}
	\label{fig: scale free no cut off}
\end{figure}

\begin{figure}[htbp] 
  	\includegraphics[scale=0.5]{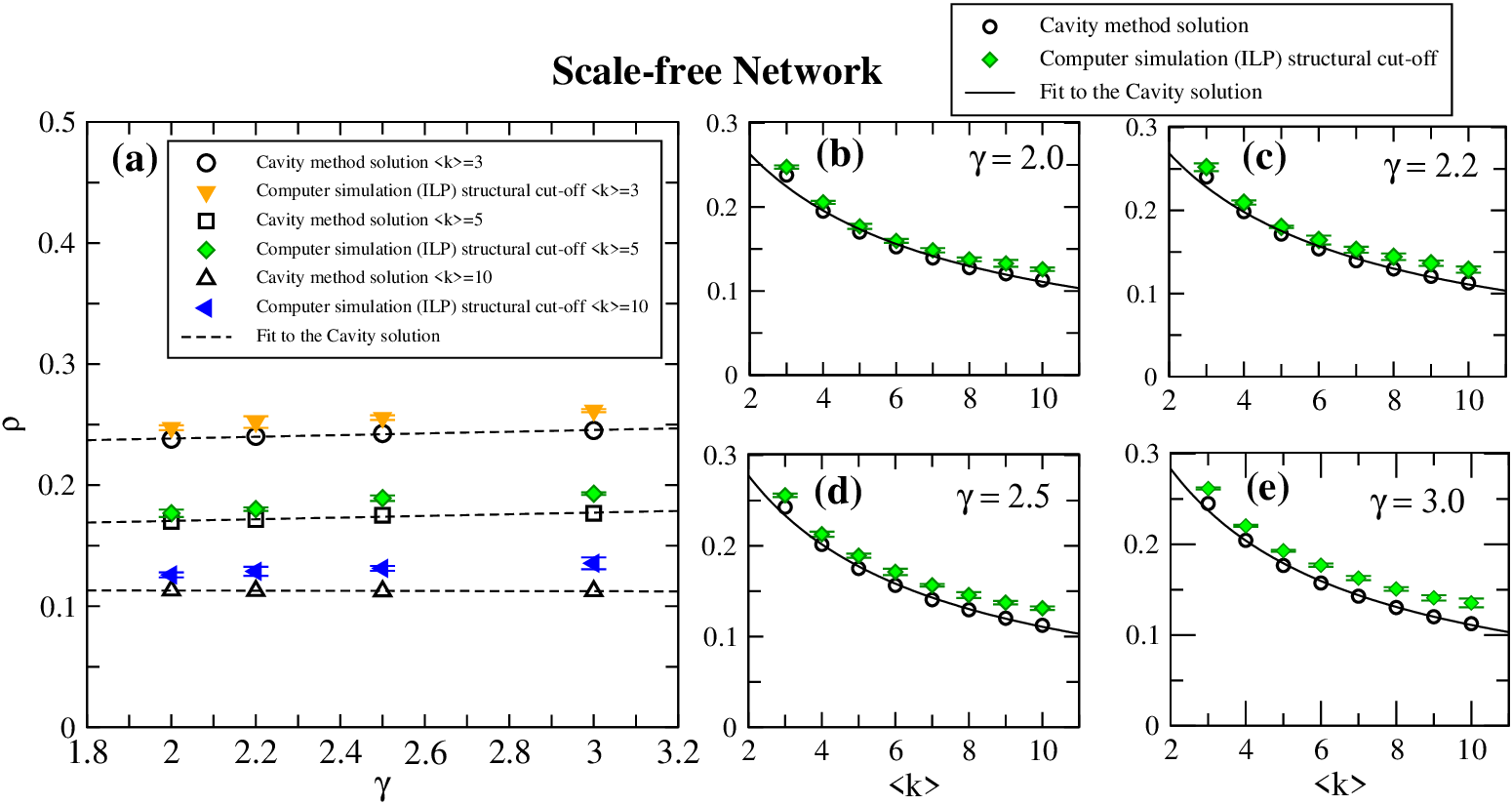}
	\caption{Comparison of the cavity method and ILP results for the MDS density $\rho$ in samples of synthetic scale-free networks generated with structural cut-off $k_c=\sqrt{<k>N}$. $\rho$ as a function of the scaling exponent $\gamma$ (a) and average degree $<k>$ (b-d) are shown in figure. The results are averaged using five network samples with a size of $N=5000$ nodes. For cavity method
the standard deviation (SD) is smaller than symbol. For ILP the SD is shown for each symbol.}
	\label{fig: scale free with cut off}
\end{figure}

Next, we examine the case of scale-free networks. We first generated samples of synthetic scale-free networks
with a variety of scaling exponent $\gamma$ and average degree $<k>$ using the Havel-Hakimi algorithm with random edge swaps (HMC). All 
samples were generated with a size of $N=5000$ nodes. We investigated two different cases. We constructed a set of scale-free 
network samples with natural cut-off $k_c=N-1$ and
another set with structural cut-off $k_c=\sqrt{<k>N}$. The minimum degree is $k_{min}=2$ in both cases. Fig. \ref{fig: scale free no cut off} 
shows the results for natural cut-off (i.e. no structural cut-off is considered). The dependence of the MDS density $\rho$ as a function of 
the average degree $<k>$ shows a good agreement with the cavity method analytical predictions (\ref{eqn: analytic formula2}), in particular 
when $\gamma$  increases (see Figs. \ref{fig: scale free no cut off} (c)-(e)). By increasing the average degree, the MDS density $\rho$ 
decreases. For small values of $\gamma$, the predictions of the cavity method deviates from the ILP solutions when average degree 
increases (see Fig. \ref{fig: scale free no cut off}(b)).  The reason is because in this case, the network tends to have hubs with 
very high degree. While these hubs are still visible to ILP method, they are not observed by the cavity method. It is well-known 
that cavity method addresses better homogeneous networks than extremely inhomogeneous networks. On the other hand, by examining 
the function of MDS density $\rho$ versus $\gamma$, we clearly observe the
influence of the average degree (see Fig. \ref{fig: scale free no cut off}(a)). Moreover, in absence of 
structural cut-off, for high average degree networks, the MDS density $\rho$ for ILP decreases faster than the solution for cavity method when $\gamma$ decreases.
\newline
\indent We have also considered the case of structural cut-off when constructing finite artificial scale-free networks 
to address the finite-size effect and eliminate degree correlations. 
The computer simulations on scale-free networks with structural cut-off shows a different picture 
(see Fig. \ref{fig: scale free with cut off}(a)-(e)). The MDS density $\rho$ does not significantly 
changes when $\gamma$ decreases and remains constant along all the range of $\gamma$ values 
(see Fig. \ref{fig: scale free with cut off}(a)). Moreover, when structural cut-off is considered, the agreement 
between cavity method and ILP results becomes more evident for any value of $\gamma$ and average degree (see Fig. \ref{fig: scale free with cut off} (b)-(e)). The reason is because the network tends to be more 
homogeneous when some hubs with extremely high-degree are knocked out by the structural cut-off.

\section{Conclusion}

Domination is not only one of the most active research areas in graph theory but also has found abundant 
real-world applications in many different fields, from engineering to social and natural sciences. With the 
recent years expansion of networks as data representation framework, domination techniques may provide a 
rich set of tools to face current network problems in society and nature. 

In this work,  by using the cavity method and the ultra-discretization procedure, we solved 
for the first time the MDS problem analytically and derived a combinatorial equation whose computation is easier than that of ILP. By 
only using the degree distribution of a network as an input information, we can compute the density of an MDS and 
investigate any dependence with respect to other network variables without using 
assistance of any complex optimization methods such as ILP or DP. 

The present analysis may allow a variety of rich extensions such as computing the 
corresponding analytical expression for MDS density in directed and bipartite networks.

\section{Acknowledgements} We thank Prof. Tatsuya Akutsu for insighful comments. J.C.N. was partially supported by MEXT, Japan (Grant-in-Aid No. 25330351), and T.O. was partially supported by JSPS Grants-in-Aid for Scientific Research (Grant Number 15K01200) and Otsuma Grant-in Aid for Individual Exploratory Research (Grant Number S2609).

\newpage

\subsection{}
\subsubsection{}

\bibliography{basename of .bib file}

\end{document}



\begin{center}
{\bf \large{Supplementary Information} }
\end{center}
\begin{center}
{\bf Analytical Solution for the Size of the Minimum Dominating Set in Complex Networks}
\end{center}
\begin{center}
{\it Jose C. Nacher and Tomoshiro Ochiai}
\end{center}
\tableofcontents

\section{   Derivation of the partition function of the Hamiltonian}

In this section, we will derive the partition function of the Hamiltonian (see Eq. (3) in main text), as in [30,42].

By removing node $a$, we define cavity partition function $Z_a$ by
\begin{eqnarray}
Z_a=\prod_{b \in \partial a}Z_{b \to a}^\prime.
\end{eqnarray}
Similarly, by removing an edge $(ab)$, we define cavity partition function $Z_{(ab)}$ by
\begin{eqnarray}
Z_{(ab)}=Z_{a \to b}^\prime Z_{b \to a}^\prime.
\end{eqnarray}
Let us define cavity free energy $\Delta f_a$ by
\begin{eqnarray}
e^{-\beta\Delta f_a} &\equiv& \frac{Z}{Z_a} \nonumber\\
&=&\sum_{\sigma_a}\sum_{\{\sigma_b | b \in \partial a\}} \sum_{\{\sigma_{b\to a} | b \in \partial a\}}
e^{-\beta\sigma_a}I_a \prod_{b \in \partial a} I_b \nu_{b \to a}(\sigma_b, \sigma_{b\to a}). \label{eqn:free enegy 1}
\end{eqnarray}
Here the last equality holds after some computation. We then define cavity free energy $\Delta f_{(ab)}$ by
\begin{eqnarray}
e^{-\beta\Delta f_{(ab)}} &\equiv& \frac{Z}{Z_{(ab)}} \nonumber\\
&=&\sum_{\sigma_a} \sum_{\sigma_{a\to b}}\sum_{\sigma_b} \sum_{\sigma_{b\to a}}
I_a I_b \nu_{a \to b}(\sigma_a, \sigma_{a\to b}) \nu_{b \to a}(\sigma_b, \sigma_{b\to a}). \label{eqn:free enegy 2}
\end{eqnarray}

Furthermore, we can transform (\ref{eqn:free enegy 1}) and  (\ref{eqn:free enegy 2}) as follows:
\begin{eqnarray}
\frac{Z}{Z_{a}} &=& e^{-\beta} + \prod_{b\in\partial a}(1-r_{00}^{b\to a}) - \prod_{b\in\partial a}r_{0}^{b\to a} \nonumber\\
&=& e^{-\beta} + \prod_{b\in\partial a}(r_0^{b\to a}+r_1^{b\to a}) - \prod_{b\in\partial a}r_{0}^{b\to a},
\end{eqnarray}
and
\begin{eqnarray}
\frac{Z}{Z_{(ab)}} &=& r_1^{a\to b} r_1^{b\to a} + r_0^{a\to b} r_0^{b\to a} + 
 r_1^{a\to b} (r_0^{b\to a} + r_{00}^{b\to a}) +  (r_0^{a\to b} + r_{00}^{a\to b}) r_1^{b\to a} \nonumber\\
&=& r_1^{a\to b} r_1^{b\to a} + r_0^{a\to b} r_0^{b\to a} + 
 r_1^{a\to b} (1 - r_1^{b\to a}) +  (1-r_1^{a\to b}) r_1^{b\to a}.
\end{eqnarray}
When the graph is a tree, we have $|V|=|E|+1$, where $|V|$ and $|E|$ is the number of nodes $V$ and edges $E$, respectively. Then, the partition function can be decomposed as follows:
\begin{eqnarray}
Z =  e^{-\beta N f} = \prod_{a \in V} \frac{Z}{Z_{a}}  \prod_{(ab) \in E} \frac{Z_{(ab)}}{Z}. 
\end{eqnarray}

By taking logarithm on the previous expression, we obtain 
\begin{eqnarray}
\log Z 
&=&  \sum_{a \in V} \log (e^{-\beta} + \prod_{b\in\partial a}(r_0^{b\to a}+r_1^{b\to a}) - \prod_{b\in\partial a}r_{0}^{b\to a}) \nonumber\\ 
&-& \sum_{(ab) \in E} \log( r_1^{a\to b} r_1^{b\to a} + r_0^{a\to b} r_0^{b\to a} + 
 r_1^{a\to b} (1 - r_1^{b\to a}) +  (1-r_1^{a\to b}) r_1^{b\to a} ). 
\end{eqnarray}
Using a coarse-grained approximation, we obtain
\begin{eqnarray}
\log Z&=& |V| \sum_{k=2}^\infty p_k\int\prod_{l=1}^k dr_1^l dr_0^l P(r_1^l,r_0^l) \nonumber\\
&&\times\log (e^{-\beta} + \prod_{l=1}^k (r_0^l+r_1^l) - \prod_{l=1}^k r_{0}^{l}) \nonumber\\
&&-|E|\int\prod_{l=1}^2 dr_1^l dr_0^l P(r_1^l,r_0^l)\log( r_1^{1} r_1^{2} + r_0^{1} r_0^{2} + 
 r_1^{1} (1 - r_1^{2}) +  (1-r_1^{1}) r_1^{2} ).
\end{eqnarray}


\section{A simple analytic expression for MDS in random networks}
Eq. (44) in main text provides us an analytical tool to compute the density of MDS for any kind of network. It may be possible, however, to obtain examine the behavior of this equation for specific network structures using some approximations. Below we derived the approximated analytical expression for the random network characterized with a Poisson degree distribution.

First, the density of the MDS can be evaluated using the following inequality:
\begin{eqnarray}
\sum_{k=2}^\infty p_{k}[1-(1-b)^k-kb(1-b)^{k-1}+a^{k}\frac{1}{k+1}] < \rho < \sum_{k=2}^\infty p_{k}[1-(1-b)^k+a^{k}\frac{1}{k+1}].
\end{eqnarray}
As a rough approximation, we set
\begin{eqnarray}
\rho = \sum_{k=0}^\infty p_{k}[1-(1-b)^k].
\end{eqnarray}
Here we start the summation from $k=0$ instead of $k=2$ as an approximation. This approximation is good if the average degree is enough large.
Let us define generating functions as follows:
\begin{eqnarray}
&&G(x)=\sum_{k=0}^\infty p_{k}x^k, \\
&&H(x)=\sum_{k=0}^\infty q_{k}x^k. 
\end{eqnarray}
Then, Eqs. (37-39) in main text can be rewritten as follows:
\begin{eqnarray}
&&a=H(1-b)-H(a), \label{eqn: H G 1}\\
&&b=H(a), \label{eqn: H G 2}\\
&&\rho = 1-G(1-b). \label{eqn: H G 3}
\end{eqnarray}

We use a Poisson distribution $p_k=\frac{e^{-z}z^k}{k!}$. Then we have $q_k=\frac{e^{-z}z^k}{k!}$. Using this distribution, we get $H(x)=G(x)=e^{z(x-1)}$.
Inserting these into (\ref{eqn: H G 1}), (\ref{eqn: H G 2}) and (\ref{eqn: H G 3}), we get
\begin{eqnarray}
&&a=e^{-zb}-e^{z(a-1)}, \label{eq: poisson1}\\
&&b=e^{z(a-1)}, \label{eq: poisson2}\\
&&\rho = 1-e^{-zb}. \label{eq: poisson3} 
\end{eqnarray}
Here we derive the simple formula for the density of MDS by using the approximation in $z\to\infty$. From Eq.(\ref{eq: poisson1}) and (\ref{eq: poisson2}), we get 
\begin{eqnarray}
1-a\sim b(z+1)\sim bz, \label{eq: poisson4} 
\end{eqnarray}
where we use the fact that $bz$ is small, since we know that $bz$ is small in numerical results. In the same way, we get
\begin{eqnarray}
\rho \sim bz. \label{eq: poisson5}
\end{eqnarray}
Inserting  Eq.(\ref{eq: poisson4}) and (\ref{eq: poisson5}) into (\ref{eq: poisson2}), we get
\begin{eqnarray}
b\sim e^{-bz^2} = e^{-\rho z} \sim 1-\rho z, \label{eq: poisson6}
\end{eqnarray}
where we use $\rho z$ is small, since we know that $\rho z$ is small in the simulated result in $z \to \infty$.
From Eq.(\ref{eq: poisson5}) and (\ref{eq: poisson6}), we obtain 
\begin{eqnarray}
\rho \sim \frac{z}{1+z^2} \sim \frac{1}{z}.
\end{eqnarray}
This result shows that a high density of links implies a low MDS density. This inverse functional dependence with average degree is also observed in cavity method and ILP solutions shown in Fig. 4b.

